\documentclass{IEEEtran}
\usepackage{cite}
\usepackage{amsmath,amssymb,amsfonts}
\usepackage{algorithmic}
\usepackage{graphicx}
\usepackage{textcomp}
\usepackage{algorithmic}
\usepackage{algorithm}
\usepackage{array}
\usepackage[caption=false,font=normalsize,labelfont=sf,textfont=sf]{subfig}
\usepackage{stfloats}
\usepackage{url}
\usepackage{verbatim}
\usepackage{enumerate}
\usepackage{tikz}
\usepackage{bm}
\usepackage{threeparttable,booktabs}
\def\BibTeX{{\rm B\kern-.05em{\sc i\kern-.025em b}\kern-.08em
    T\kern-.1667em\lower.7ex\hbox{E}\kern-.125emX}}
\begin{document}
\title{Generalized Scattering Matrix of Antenna: Moment Solution, Compression Storage and Application}

\author{Chenbo Shi, Jin Pan, Xin Gu, Shichen Liang and Le Zuo
\thanks{Received Aug. 16, 2024. Revised Mar. 13, 2025. This work was supported by the Aeronautical Science Fund under Grant ASFC-20220005080001. (\textit{Corresponding author: Jin Pan.})}
\thanks{Chenbo Shi, Jin Pan, Xin Gu and Shichen Liang are with the School of Electronic Science and Engineering, University of Electronic Science and Technology of China, Chengdu 611731 China  (e-mail: chenbo\_shi@163.com; panjin@uestc.edu.cn; xin\_gu04@163.com; lscstu001@163.com).}
\thanks{Le Zuo is with The 29th Research Institute of China Electronics Technology Group Corporation (e-mail: zorro1204@163.com)}
}

\markboth{ }%
{Shell \MakeLowercase{\textit{et al.}}: A Sample Article Using IEEEtran.cls for IEEE Journals}

\maketitle

\begin{abstract}
This paper presents a computation method of generalized scattering matrix (GSM) based on integral equations and the method of moments (MoM), specifically designed for antennas excited through waveguide ports. By leveraging two distinct formulations---magnetic-type and electric-type integral equations---we establish concise algebraic relations linking the GSM directly to the impedance matrices obtained from MoM. To address practical challenges in storing GSM data across wide frequency bands and multiple antenna scenarios, we propose a efficient compression scheme. This approach alleviates memory demands by selectively storing the dominant eigencomponents that govern scattering behavior. Numerical validation examples confirm the accuracy of our method by comparisons with full-wave simulation results. Furthermore, we introduce an efficient iterative procedure to predict antenna array performance, highlighting remarkable improvements in computational speed compared to conventional numerical methods. These results collectively demonstrate the GSM framework's strong potential for antenna-array design processes.
\end{abstract}

\begin{IEEEkeywords}
  Generalized scattering matrix, T-matrix, the method of moments, integral equations, theory of characteristic modes.
\end{IEEEkeywords}

\section{Introduction}

\IEEEPARstart{T}{he} generalized scattering matrix (GSM) characterizes the radiation and scattering properties of antennas, serving as a fundamental tool for the analysis and design of minimal scattering antennas \cite{ref_MS_antenna_Smat1,ref_MS_antenna_Smat2}. By employing the translation theorem of spherical wavefunctions \cite{ref_scattering_theory,ref_sph_addition1,ref_sph_addition2}, the GSM provides an analytical representation of mutual coupling among antennas. Such applications frequently appear in scenarios involving mutual coupling, such as spherical near-field measurements \cite{ref_sph_near_measure}. This method also enables accurate predictions of antenna array behavior based on independent simulations of individual elements.

Despite these advantages, computational research on the GSM remains limited. One existing approach adopts a finite element method (FEM) framework \cite{ref_Smat_FEM}, where a spherical truncation box (air box) surrounds the antenna at a prescribed radius, and radiation boundary conditions are enforced through the method of unimoment \cite{ref_KK_MEI} (not a mainstream way to process the radiation boundary in FEM). Although FEM is effective for complex geometries and inhomogeneous media, it may be constrained in some applications, such as large-scale problems, as it requires discretization of the entire surrounding air box. Additionally, as pointed out by \cite{ref_Smat_FEM}, the necessity of using a spherical truncation box reduces its efficiency for elongated structures.

Another method uses characteristic mode theory to compute the GSM \cite{ref_Smat_CMA}, linking the weighting coefficients of characteristic modes directly to spherical wavefunction expansion coefficients. This connection facilitates obtaining each GSM block through superpositions of characteristic modes. While integral equation-based characteristic mode decomposition circumvents certain FEM limitations, challenges arise regarding series truncation---characteristic mode expansions typically exhibit slow convergence near antenna feeding ports, except for relatively simple structures like half-wave dipoles \cite{ref_CM_unify,ref_CM_souce_exp_converge1,ref_CM_souce_exp_converge2}. Furthermore, the reliance on ideal lumped ports, supporting only a single propagating mode, narrows the scope of realistic antenna applications.

Motivated by these shortcomings, this paper proposes an integral-equation method leveraging the method of moments (MoM) framework to rigorously determine the GSM of waveguide-fed antennas. This draws inspiration from the celebrated T-matrix formalism commonly employed in electromagnetic scattering analyses \cite{ref_Waterman}. Early pioneering work by \cite{ref_Kim} employed MoM to derive the T-matrix of metallic scatterers through spherical wave expansions of the dyadic Green's function. A more recent refinement by \cite{ref_hybrid} introduced an elegant, simplified symbolic representation using real-valued vector spherical wavefunctions, preserving the essence of \cite{ref_Kim}'s foundational work. We adopt selected notational conventions from \cite{ref_hybrid}, but extend its formulation to incorporate magnetic current contributions arising from waveports. Moreover, Unlike pure scatterers, antennas fed through waveports inherently lack a straightforward T-matrix definition, necessitating a novel direct algebraic linkage between the GSM and the impedance matrix intrinsic to MoM.

To achieve this objective, we first establish two types of integral equations accounting for waveport excitation. Subsequently, we derive a clear and concise algebraic relation between the GSM and impedance matrix using a projection operator, bridging antenna-surface currents with vector spherical and waveguide modal expansions. This representation uncovers an intrinsic mathematical connection: the eigenvalues of GSM naturally correspond to the generalized eigenvalues (characteristic numbers) of the impedance matrix. Exploiting this powerful link, we introduce efficient data-compression strategies: For multi-antenna and broadband applications, the GSM's eigenvalue decomposition reduces its storage demands, whereas singular value decomposition emerges as an alternative in scenarios involving loss.

Moreover, we develop an iterative method harnessing the GSM to solve antenna array problems. In numerical validation involving a 20-element dipole array, our method achieved computational speeds multiple times greater than those offered by the multilayer fast multipole method (MLFMM). This remarkable performance demonstrates the GSM's considerable promise for accelerated yet precise antenna array simulations.

\section{Integral Equations for Wave-port Feeding}
\label{SecII}
As a foundational element of this work, we first introduce the MoM formulation for electromagnetic problems involving waveports. Although this approach traces back to \cite{ref_waveguide_Junction}, it remains relatively underutilized in contemporary research. More recent developments can be found in \cite{ref_waveguide_MoM1,ref_waveguide_MoM2}. Here, we adopt a distinct notation system, which, as we will demonstrate, establishes an explicit algebraic connection between the moment matrices and the GSM. This connection provides significant theoretical advantages, particularly in understanding the eigenvalue problem.

\begin{figure}[!t]
  \centering
  \includegraphics[]{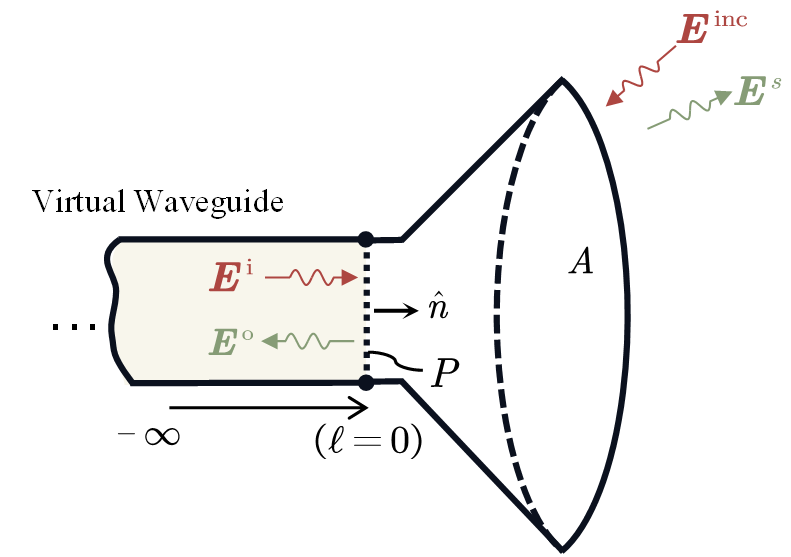}
  \caption{Field representations of an antenna fed by a waveguide. The surface P represents the waveport, which is assumed to connect seamlessly to a perfectly matched, semi-infinite waveguide.}
  \label{fWaveport}
\end{figure}

Figure \ref{fWaveport} illustrates a typical waveport-fed antenna model, where P represents the waveport surface and A denotes the antenna surface. The waveport is modeled as a reflection-free boundary, conceptually linked to a semi-infinite waveguide filled with a homogeneous medium. Equivalent electric currents $\boldsymbol{J}^e_\mathrm{P}$ and magnetic current $\boldsymbol{J}^m$ reside on the waveport surface. For simplicity, we assume that the antenna is a perfect electric conductor (PEC), thereby restricting its current distribution to electric currents $\boldsymbol{J}^e_\mathrm{A}$.

The total electromagnetic field external to the waveguide consists of incident and scattered components:
\begin{equation}
  \boldsymbol{E}^{\mathrm{ext}}=\boldsymbol{E}^{\mathrm{inc}}+\boldsymbol{E}^s,\boldsymbol{H}^{\mathrm{ext}}=\boldsymbol{H}^{\mathrm{inc}}+\boldsymbol{H}^s.
\end{equation}
Here, the scattered fields $\boldsymbol{E}^s,\boldsymbol{H}^s$ arise due to the radiation of $\boldsymbol{J}_{\mathrm{A}}^{e}$, $\boldsymbol{J}_{\mathrm{P}}^{e}$ and $\boldsymbol{J}^m$:
\begin{equation}
  \label{eq2}
  \begin{split}
    &\boldsymbol{E}^s\left( \boldsymbol{r} \right) =-\mathrm{j}k\eta _0\mathcal{L} \left( \boldsymbol{J}^e \right) -\mathcal{K} \left( \boldsymbol{J}^m \right) \\
    &\boldsymbol{H}^s\left( \boldsymbol{r} \right) =\mathcal{K} \left( \boldsymbol{J}^e \right) -\mathrm{j}k/\eta _0\mathcal{L} \left( \boldsymbol{J}^m \right)
  \end{split}
\end{equation}
where $\boldsymbol{J}^e=\boldsymbol{J}_{\mathrm{A}}^{e}\cup \boldsymbol{J}_{\mathrm{P}}^{e}$. The operators $\mathcal{L}$ and $\mathcal{K}$ are defined as
\begin{equation}
  \mathcal{L} \left( \boldsymbol{X} \right) \left\{ \boldsymbol{r} \right\} =\left< \bar{\mathbf{G}}_e\left( \boldsymbol{r},\boldsymbol{r}^\prime \right) ,\boldsymbol{X}\left( \boldsymbol{r}^{\prime} \right) \right> _S,\mathcal{K} =\nabla \times \mathcal{L}.
\end{equation}
Here, the symmetric product is given by $\left< \boldsymbol{X},\boldsymbol{Y} \right> _S=\int_S{\boldsymbol{X}\left( \boldsymbol{r}^{\prime} \right) \cdot \boldsymbol{Y}\left( \boldsymbol{r}^{\prime} \right) \mathrm{d}S}$, and $\bar{\mathbf{G}}_e\left( \boldsymbol{r},\boldsymbol{r}^\prime \right) $ represents the free-space dyadic Green's function.

Within the semi-infinite waveguide, the total field comprises both forward-propagating (incoming) and backward-propagating (outgoing) waves:
\begin{equation}
  \label{eq4}
  \boldsymbol{E}^{\mathrm{int}}=\boldsymbol{E}^i+\boldsymbol{E}^o,\boldsymbol{H}^{\mathrm{int}}=\boldsymbol{H}^i+\boldsymbol{H}^o.
\end{equation}
By enforcing the continuity of the tangential fields across the waveport surface P, we establish the equations: 
\begin{equation}
  \label{eq5}
  \begin{cases}
    \hat{n}\times \left( \boldsymbol{E}^{\mathrm{ext}}-\boldsymbol{E}^{\mathrm{int}} \right) =0\\
    \hat{n}\times \left( \boldsymbol{H}^{\mathrm{ext}}-\boldsymbol{H}^{\mathrm{int}} \right) =0\\
  \end{cases},\boldsymbol{r}\in P
\end{equation}
Meanwhile, on the antenna surface A, the PEC boundary condition is imposed: 
\begin{equation}
  \hat{n}\times \boldsymbol{E}^{\mathrm{ext}}=0,\boldsymbol{r}\in A.
\end{equation}

To solve for $\boldsymbol{E}^o$ and $\boldsymbol{H}^o$ in \eqref{eq4}, which are generated by the radiation of $\boldsymbol{J}_{\mathrm{A}}^{e}$, $\boldsymbol{J}_{\mathrm{P}}^{e}$ and $\boldsymbol{J}^m$, we invoke Schelkunoff's equivalence principle. By leveraging field continuity on both sides of P, we introduce an equivalent strategy: setting the equivalent magnetic current $-\boldsymbol{J}^m$ and a PEC wall, or alternatively, imposing an equivalent electric current $-\boldsymbol{J}_{\mathrm{A}}^{e}$ and a perfect magnetic conductor (PMC) wall. While prior works predominantly employ a PEC wall \cite{ref_waveguide_Junction,ref_waveguide_MoM1,ref_waveguide_MoM2}, we present both configurations. Recognizing their equivalence in determining the final solution, however, adopting the PMC wall can establish a direct connection between the eigenvalues of the MoM impedance matrix and the GSM---a crucial insight explored further in Sec. \ref{Sec_IV}.

\subsection{Magnetic-type Equations---Magnetic Currents and PEC Boundary}

The transverse components of the forward-propagating wave, $\boldsymbol{E}^i ,\boldsymbol{H}^i$, at a given position $\ell$ can be expressed as 
\begin{equation}
  \begin{split}
    &\boldsymbol{E}^i\left( \ell \right) =\sum_{\alpha}{\sqrt{\eta _{\alpha}}v_{\alpha}\mathbf{e}_{\alpha}e^{-\mathrm{j}\beta_\alpha \ell}}\\
    &\boldsymbol{H}^i\left( \ell \right) =\sum_{\alpha}{\frac{1}{\sqrt{\eta _{\alpha}}}v_{\alpha}\hat{n}\times \mathbf{e}_{\alpha}e^{-\mathrm{j}\beta_\alpha \ell}}.
  \end{split}
\end{equation}
due to the assumption of a homogeneous waveguide medium. This equation is carefully normalized to ensure that the incoming power is only related to the modal coefficient $v_\alpha$. Here, $\mathbf{e}_{\alpha}$ denote the transverse electric field components of the $\alpha$-th eigenmode. $\eta_{\alpha}$ and $\beta_\alpha$ correspond to the mode-specific wave impedance and propagation constant.

Similarly, the transverse components of the backward-propagating wave, $\boldsymbol{E}^o$ and $\boldsymbol{H}^o$, take the form:
\begin{equation}
  \label{eq8}
  \begin{split}
    &\boldsymbol{E}^o\left( \ell \right) =\sum_{\alpha}{\sqrt{\eta _{\alpha}}w_{\alpha}\mathbf{e}_{\alpha}e^{\mathrm{j}\beta_\alpha \ell}}\\
    &\boldsymbol{H}^o\left( \ell \right) =-\sum_{\alpha}{\frac{1}{\sqrt{\eta _{\alpha}}}w_{\alpha}\hat{n}\times \mathbf{e}_{\alpha}e^{\mathrm{j}\beta_\alpha \ell}}.
  \end{split}
\end{equation}

On the waveport surface P, the equivalent magnetic current follows from the boundary condition:
\begin{equation}
  \label{eq9}
  -\boldsymbol{J}^m=\hat{n}\times \boldsymbol{E}^{\mathrm{int}}|_{\ell =0}.
\end{equation}
Substituting the modal expansion of $ \boldsymbol{E}^{\mathrm{int}}$ leads to: 
\begin{equation}
  \label{eq10}
  -\boldsymbol{J}^m=\sum_{\alpha}{\sqrt{\eta _{\alpha}}\hat{n}\times \mathbf{e}_{\alpha}\left( v_{\alpha}+w_{\alpha} \right)}.
\end{equation}
By taking the inner product with $\hat{n} \times \mathbf{e}_{\gamma}$ and integrating over P, we obtain:
\begin{equation}
  \label{eq11}
  v_{\alpha}+w_{\alpha}=-\frac{\left< \hat{n}\times \mathbf{e}_{\alpha},\boldsymbol{J}^m \right>}{\sqrt{\eta _{\alpha}}}
\end{equation}
where the orthogonality condition $\left< \mathbf{e}_{\alpha}, \mathbf{e}_{\gamma} \right>_{\mathrm{P}} = \delta_{\alpha \gamma}$ has been applied, with $\delta_{\alpha\gamma}$ being the Kronecker delta.

Replacing $w_\alpha$ in the backward-wave component of $\boldsymbol{H}^{\mathrm{int}}$ using \eqref{eq11}, we have
\begin{equation}
  \boldsymbol{H}^{\mathrm{int}}|_{\ell =0}=\sum_{\alpha}{\frac{2}{\sqrt{\eta _{\alpha}}}v_{\alpha}\hat{n}\times \mathbf{e}_{\alpha}}+\sum_{\alpha}{\frac{\left< \hat{n}\times \mathbf{e}_{\alpha},\boldsymbol{J}^m \right>}{\eta _{\alpha}}\hat{n}\times \mathbf{e}_{\alpha}}.
\end{equation}
Substituting this expression, along with \eqref{eq9}, into the boundary conditions in \eqref{eq5}, and incorporating \eqref{eq2}, we establish the integral equations:
\begin{equation}
  \label{eq13}
  \begin{split}
    &\mathrm{j}k\eta _0\mathcal{L} \left( \boldsymbol{J}_{A}^{e} \right) +\mathrm{j}k\eta _0\mathcal{L} \left( \boldsymbol{J}_{P}^{e} \right) +\mathcal{K} \left( \boldsymbol{J}^m \right) =\boldsymbol{E}^{\mathrm{inc}},\boldsymbol{r}\in A\\
    &\mathrm{j}k\eta _0\mathcal{L} \left( \boldsymbol{J}_{A}^{e} \right) +\mathrm{j}k\eta _0\mathcal{L} \left( \boldsymbol{J}_{P}^{e} \right) +\mathcal{K}^+ \left( \boldsymbol{J}^m \right) =\boldsymbol{E}^{\mathrm{inc}},\boldsymbol{r}\in P\\
    &-\mathcal{K}^- \left( \boldsymbol{J}_{A}^{e} \right) -\mathcal{K}^- \left( \boldsymbol{J}_{P}^{e} \right) +\mathrm{j}k/\eta _0\mathcal{L} \left( \boldsymbol{J}^m \right)\\
    &+\sum_{\alpha}{\frac{\left< \hat{n}\times \mathbf{e}_{\alpha},\boldsymbol{J}^m \right>}{\eta _{\alpha}}\hat{n}\times \mathbf{e}_{\alpha}}=\boldsymbol{H}^{\mathrm{inc}}-\sum_{\alpha}{\frac{2\hat{n}\times \mathbf{e}_{\alpha}}{\sqrt{\eta _{\alpha}}}v_{\alpha}}.
  \end{split}
\end{equation}
Here, the operator $\mathcal{K} ^{\pm}\left( \boldsymbol{X} \right) =\pm \frac{1}{2}\hat{n}\times \boldsymbol{X}+\text{P.V.}\mathcal{K} \left( \boldsymbol{X} \right) $, where the first term vanishes for $\boldsymbol{r}\ne\boldsymbol{r}'$. For penetrable antenna materials, the contributions from magnetic currents on surface A must be included, and the first equation in \eqref{eq13} is replaced by the Poggio-Miller-Chang-Harrington-Wu (PMCHW) formulation \cite{ref_PMCHW}.

To discretize \eqref{eq13} using MoM, we expand $\boldsymbol{J}^e=\boldsymbol{J}^e_\mathrm{A}\cup\boldsymbol{J}^e_\mathrm{P}$ and $\boldsymbol{J}^m$ through RWG basis functions $\left\{ \boldsymbol{\psi}_i \right\}$:
\begin{equation}
  \boldsymbol{J}^e=\sum_i{I_{i}^{e}\boldsymbol{\psi }_i,\boldsymbol{J}^m}=\sum_i{I_{i}^{m}\boldsymbol{\psi }_i}.
\end{equation}
Galerkin testing then produces the algebraic equation:
\begin{equation}
  \mathbf{Z}^{\mathrm{M}}\mathbf{I}=\mathbf{V}^{\mathrm{inc}}+\mathbf{V}^{\mathrm{M}}
\end{equation}
where
\begin{equation}
  \label{eq16}
  \begin{split}
    &\mathbf{Z}^{\mathrm{M}}=\begin{bmatrix}
      \mathrm{j}k\eta _0\mathbf{L}&		-\mathrm{j}\mathbf{K}^+\\
      -\mathrm{j}\mathbf{K}^-&		\mathrm{j}k/\eta _0\mathbf{L}+\mathbf{G}^{\mathrm{M}}\\
    \end{bmatrix}
    \\
   & \mathbf{I}=\begin{bmatrix}
      \mathbf{I}^e\\
      \mathrm{j}\mathbf{I}^m\\
    \end{bmatrix} ,\mathbf{V}^{\mathrm{inc}}=\begin{bmatrix}
      \mathbf{V}^e\\
      \mathrm{j}\mathbf{V}^m\\
    \end{bmatrix} =\begin{bmatrix}
      \left< \left\{ \boldsymbol{\psi }_i \right\} ,\boldsymbol{E}^{\mathrm{inc}} \right> _{\mathrm{A}}\\
      \mathrm{j}\left< \left\{ \boldsymbol{\psi }_i \right\} ,\boldsymbol{H}^{\mathrm{inc}} \right> _{\mathrm{P}}\\
    \end{bmatrix}
    \\
    &\mathbf{V}^{\mathrm{M}}=\begin{bmatrix}
      \mathbf{O}\\
      \mathrm{j}\mathbf{V}_{\mathrm{P}}^{m}\\
    \end{bmatrix} ,\left[ \mathbf{V}_{\mathrm{P}}^{m} \right] _i=-2\left< \boldsymbol{\psi }_i,\sum_{\alpha}{v_{\alpha}\frac{\hat{n}\times \mathbf{e}_{\alpha}}{\sqrt{\eta _{\alpha}}}} \right> _{\mathrm{P}}
  \end{split}
\end{equation}
and
\begin{equation}
  \label{eq15}
  \begin{split}
    &\left[ \mathbf{L} \right] _{ij}=\left< \boldsymbol{\psi }_i,\mathcal{L} \left( \boldsymbol{\psi }_j \right) \right> ,\left[ \mathbf{K}^{\pm} \right] _{ij}=\left< \boldsymbol{\psi }_i,\mathcal{K} ^{\pm}\left( \boldsymbol{\psi }_j \right) \right> \\
    &\left[ \mathbf{G}^{\mathrm{M}} \right] _{ij}=\frac{1}{\eta _{\alpha}}\sum_{\alpha}{\left< \boldsymbol{\psi }_i,\hat{n}\times \mathbf{e}_{\alpha} \right> _{\mathrm{P}}\left< \boldsymbol{\psi }_j,\hat{n}\times \mathbf{e}_{\alpha} \right> _{\mathrm{P}}}.
  \end{split}
\end{equation}
Note that the summation series defining $\mathbf{G}^{\mathrm{M}}$ and $\mathbf{V}^{\mathrm{M}}$ can be truncated after $M$ terms, encompassing all propagating TEM, TE, and TM modes along with several lower-order evanescent modes.

Solving the linear system \eqref{eq15} for the magnetic current coefficients $\mathbf{I}^m$ in $\mathbf{I}$,  and subsequently substituting into \eqref{eq11}, yields the reflected wave coefficients $w_{\alpha}$. Alternatively, we introduce a projection operator $\mathbf{Q}^{\mathrm{M}}=\begin{bmatrix}	\mathbf{O}&		\tilde{\mathbf{Q}}^{\mathrm{M}}\end{bmatrix}$ with the sub-block defined explicitly as:
\begin{equation}
  \left[ \tilde{\mathbf{Q}}^{\mathrm{M}} \right] _{\alpha i}=-\frac{\mathrm{j}}{\sqrt{\eta _{\alpha}}}\left< \hat{n}\times \mathbf{e}_{\alpha},\boldsymbol{\psi }_i \right> _{\mathrm{P}}.
\end{equation}
Equation \eqref{eq11} thus takes a concise vectorized form:
\begin{equation}
  \label{eq19}
  \mathbf{v}+\mathbf{w}=-\mathbf{Q}^{\mathrm{M}}\mathbf{I}.
\end{equation}
In addition, the matrix identity: $\begin{bmatrix}
	\mathbf{O}&		\mathbf{O}\\
	\mathbf{O}&		\mathbf{G}^{\mathrm{M}}\\
\end{bmatrix} =-\mathbf{Q}^{\mathrm{M}t}\mathbf{Q}^{\mathrm{M}}$ with $(\cdot)^t$ denoting matrix transpose, further streamlines computational implementations.

\subsection{Electric-type Equations---Electrical Currents and PMC Boundary}

For the electric-type integral equations, we impose boundary conditions based on the magnetic fields at the waveport surface. Thus, the equivalent electric current on surface P is expressed as:
\begin{equation}
  \label{eq20}
  \boldsymbol{J}_{\mathrm{P}}^{e}=\hat{n}\times \boldsymbol{H}^{\mathrm{int}}|_{\ell =0}.
\end{equation}
Using modal expansions, this relation becomes
\begin{equation}
  \label{eq21}
  \boldsymbol{J}_{\mathrm{P}}^{e}=\hat{n}\times \sum_{\alpha}{\frac{1}{\sqrt{\eta _{\alpha}}}\left( v_{\alpha}-w_{\alpha} \right) \hat{n}\times \mathbf{e}_{\alpha}}.
\end{equation}
Taking the inner product of $\mathbf{e}_{\gamma}$ on \eqref{eq21}, integrating over surface P, and invoking orthogonality yields: 
\begin{equation}
  \label{eq22}
  w_{\alpha}-v_{\alpha}=\sqrt{\eta _{\alpha}}\left< \mathbf{e}_{\alpha},\boldsymbol{J}_{\mathrm{P}}^{e} \right> 
\end{equation}
then expressing the internal electric field at the port ($\ell=0$) using this relation gives:
\begin{equation}
  \label{eq23}
  \boldsymbol{E}^{\mathrm{int}}|_{\ell =0}=\sum_{\alpha}{2\sqrt{\eta _{\alpha}}v_{\alpha}\mathbf{e}_{\alpha}}+\sum_{\alpha}{\eta _{\alpha}\left< \mathbf{e}_{\alpha},\boldsymbol{J}^e \right> \mathbf{e}_{\alpha}}.
\end{equation}

Substituting \eqref{eq23} and \eqref{eq20} into the boundary conditions \eqref{eq5}, and employing \eqref{eq2} yields the final form of the integral equations: 
\begin{equation}
  \label{eq24}
  \begin{split}
   & \mathrm{j}k\eta _0\mathcal{L} \left( \boldsymbol{J}_{A}^{e} \right) +\mathrm{j}k\eta _0\mathcal{L} \left( \boldsymbol{J}_{P}^{e} \right) +\mathcal{K} \left( \boldsymbol{J}^m \right) =\boldsymbol{E}^{\mathrm{inc}},\boldsymbol{r}\in A\\
   &\mathrm{j}k\eta _0\mathcal{L} \left( \boldsymbol{J}_{A}^{e} \right) +\mathrm{j}k\eta _0\mathcal{L} \left( \boldsymbol{J}_{P}^{e} \right)+\sum_{\alpha}{\eta _{\alpha}\left< \mathbf{e}_{\alpha},\boldsymbol{J}_{\mathrm{P}}^{e} \right> \mathbf{e}_{\alpha}}\\
   & \ \quad\qquad +\mathcal{K}^- \left( \boldsymbol{J}^m \right) =\boldsymbol{E}^{\mathrm{inc}}-\sum_{\alpha}{2\sqrt{\eta _{\alpha}}v_{\alpha}\mathbf{e}_{\alpha}},\boldsymbol{r}\in P\\
   & -\mathcal{K} \left( \boldsymbol{J}_{A}^{e} \right) -\mathcal{K}^+ \left( \boldsymbol{J}_{P}^{e} \right) +\mathrm{j}k/\eta _0\mathcal{L} \left( \boldsymbol{J}^m \right) =\boldsymbol{H}^{\mathrm{inc}},\boldsymbol{r}\in P.
  \end{split}
\end{equation}

Employing the same MoM discretization approach as described previously, equation \eqref{eq24} is cast into the matrix form: 
\begin{equation}
  \mathbf{Z}^{\mathrm{E}}\mathbf{I}=\mathbf{V}^{\mathrm{inc}}+\mathbf{V}^{\mathrm{E}}
\end{equation}
where
\begin{equation}
  \begin{split}
    &\mathbf{Z}^{\mathrm{E}}=\begin{bmatrix}
      \mathrm{j}k\eta _0\mathbf{L}+\mathbf{G}^{\mathrm{E}}&		-\mathrm{j}\mathbf{K}^-\\
      -\mathrm{j}\mathbf{K}^+&		\mathrm{j}k/\eta _0\mathbf{L}\\
    \end{bmatrix}
    \\
    &\left[ \mathbf{G}^{\mathrm{E}} \right] _{ij}=\eta _{\alpha}\sum_{\alpha}{\left< \boldsymbol{\psi }_i,\mathbf{e}_{\alpha} \right> _{\mathrm{P}}\left< \boldsymbol{\psi }_j,\mathbf{e}_{\alpha} \right> _{\mathrm{P}}}
    \\
   & \mathbf{V}^{\mathrm{E}}=\begin{bmatrix}
      \mathbf{V}_{\mathrm{P}}^{e}\\
      \mathbf{O}\\
    \end{bmatrix} ,\left[ \mathbf{V}_{\mathrm{P}}^{e} \right] _i=-2\left< \boldsymbol{\psi }_i,\sum_{\alpha}{v_{\alpha}\sqrt{\eta _{\alpha}}\mathbf{e}_{\alpha}} \right> _{\mathrm{P}}
  \end{split}
\end{equation}

To facilitate concise representation, we introduce a projection operator $\mathbf{Q}^{\mathrm{E}}=\begin{bmatrix}
	\tilde{\mathbf{Q}}^{\mathrm{E}}&		\mathbf{O}\\
\end{bmatrix}$, with the elements defined explicitly as
\begin{equation}
  \left[ \tilde{\mathbf{Q}}^{\mathrm{E}} \right] _{\alpha i}=-\sqrt{\eta _{\alpha}}\left< \mathbf{e}_{\alpha},\boldsymbol{\psi }_i \right> _{\mathrm{P}}.
\end{equation}
Thus, equation \eqref{eq22} can be succinctly vectorized as: 
\begin{equation}
  \label{eq28}
  \mathbf{w}-\mathbf{v}=-\mathbf{Q}^{\mathrm{E}}\mathbf{I}.
\end{equation}

Additionally, the identity $\begin{bmatrix}
	\mathbf{G}^{\mathrm{E}}&		\mathbf{O}\\
	\mathbf{O}&		\mathbf{O}\\
\end{bmatrix} =\mathbf{Q}^{\mathrm{E}t}\mathbf{Q}^{\mathrm{E}}$ provides further computational simplification.

\section{Generalized Scattering Matrix and MoM Evaluation}
\subsection{Expanding Fields into Vector Spherical Wavefunctions}

To define the GSM of antennas, the external electromagnetic fields $\boldsymbol{E}^{\mathrm{ext}}, \boldsymbol{H}^{\mathrm{ext}}$ must be expanded into incoming and outgoing spherical vector waves \cite{ref_scattering_Montgomery,ref_scattering_expansion,ref_sph_near_measure}:
\begin{equation}
  \label{eq29}
  \begin{split}
    &\boldsymbol{E}^{\mathrm{ext}}\left( \boldsymbol{r} \right) =k\sqrt{\eta _0}\sum_{\alpha}{\left[ a_{\alpha}\mathbf{u}_{\alpha}^{\left( 3 \right)}\left( k\boldsymbol{r} \right) +b_{\alpha}\mathbf{u}_{\alpha}^{\left( 4 \right)}\left( k\boldsymbol{r} \right) \right]}\\
    &\boldsymbol{H}^{\mathrm{ext}}\left( \boldsymbol{r} \right) =\mathrm{j}k/\sqrt{\eta _0}\sum_{\alpha}{\left[ a_{\alpha}\mathbf{u}_{\bar{\alpha}}^{\left( 3 \right)}\left( k\boldsymbol{r} \right) +b_{\alpha}\mathbf{u}_{\bar{\alpha}}^{\left( 4 \right)}\left( k\boldsymbol{r} \right) \right]}
  \end{split}
\end{equation}
Here, $\mathbf{u}_{\alpha}^{\left( p \right)}\left( k\boldsymbol{r} \right)$ are vector spherical wavefunctions \cite{ref_hybrid}, with superscripts $p=3$ and $p=4$ indicating incoming and outgoing waves, respectively. We also introduce regular spherical waves $\mathbf{u}_{\alpha}^{\left( 1 \right)}\left( k\boldsymbol{r} \right)$, defined as the real components of $\mathbf{u}_{\alpha}^{\left( 3,4 \right)}\left( k\boldsymbol{r} \right)$. The combined mode index $\alpha \rightarrow \tau \sigma lm$ incorporates polarization type $\tau =\left\{ 1,2 \right\}$ (TE, TM), parity $\sigma =\left\{ e,o \right\}$ (even, odd), angular degree $l=\left\{ 1,2,\cdots L_{\max} \right\}$, and order $m=\left\{ 0,1,\cdots ,l \right\}$. Additionally, $\bar{\alpha}=\bar{\tau}\sigma lm$, with $\bar{\tau}=3-\tau$ interchanges TE and TM types, yielding the relation $\mathbf{u}_{\bar{\alpha}}^{\left( p \right)}=k\nabla \times \mathbf{u}_{\alpha}^{\left( p \right)}$. The maximum degree $L_{\max}$ required for numerical accuracy is estimated by \cite{ref_Sph_deg_trunction}: 
\begin{equation}
  L_{\max}=\lceil kr_{\min}+7\sqrt[3]{kr_{\min}}+3 \rceil 
\end{equation}
where $r_{\min}$ is the minimal radius of a sphere enclosing the antenna, resulting in $J=2L_{\max}\left( L_{\max}+2 \right)$ vector spherical waves.

Using the identity $\mathbf{u}_{\alpha}^{\left( 3 \right)} = 2\mathbf{u}_{\alpha}^{\left( 1 \right)} - \mathbf{u}_{\alpha}^{\left( 4 \right)}$, equation \eqref{eq29} can alternatively be expressed as
\begin{equation}
  \begin{split}
    &\boldsymbol{E}^{\mathrm{ext}}\left( \boldsymbol{r} \right) =k\sqrt{\eta _0}\sum_{\alpha}{\left[ g_{\alpha}\mathbf{u}_{\alpha}^{\left( 1 \right)}\left( k\boldsymbol{r} \right) +h_{\alpha}\mathbf{u}_{\alpha}^{\left( 4 \right)}\left( k\boldsymbol{r} \right) \right]}\\
    &\boldsymbol{H}^{\mathrm{ext}}\left( \boldsymbol{r} \right) =\mathrm{j}k/\sqrt{\eta _0}\sum_{\alpha}{\left[ g_{\alpha}\mathbf{u}_{\bar{\alpha}}^{\left( 1 \right)}\left( k\boldsymbol{r} \right) +h_{\alpha}\mathbf{u}_{\bar{\alpha}}^{\left( 4 \right)}\left( k\boldsymbol{r} \right) \right]}.
  \end{split}
\end{equation}
This separates the total external field into incident $\boldsymbol{E}^{\mathrm{inc}}, \boldsymbol{H}^{\mathrm{inc}}$ and scattered fields $\boldsymbol{E}^s, \boldsymbol{H}^s$:
\begin{equation}
  \label{eq32}
  \begin{split}
    &\boldsymbol{E}^{\mathrm{inc}}=k\sqrt{\eta _0}\sum_{\alpha}{g_{\alpha}\mathbf{u}_{\alpha}^{\left( 1 \right)}},\boldsymbol{E}^s=k\sqrt{\eta _0}\sum_{\alpha}{h_{\alpha}\mathbf{u}_{\alpha}^{\left( 4 \right)}}\\
    &\boldsymbol{H}^{\mathrm{inc}}=\mathrm{j}k/\sqrt{\eta _0}\sum_{\alpha}{g_{\alpha}\mathbf{u}_{\bar{\alpha}}^{\left( 1 \right)}},\boldsymbol{H}^s=\mathrm{j}k/\sqrt{\eta _0}\sum_{\alpha}{h_{\alpha}\mathbf{u}_{\bar{\alpha}}^{\left( 4 \right)}}
  \end{split}
\end{equation}
where
\begin{equation}
  \label{eq33}
  \mathbf{g}=2\mathbf{a},\mathbf{h}=\mathbf{b}-\mathbf{a}.
\end{equation}

Since the free-space dyadic Green's function admits a spherical-wave expansion of the form \cite{ref_scattering_theory,ref_Kim,ref_hybrid}
\begin{equation}
  \label{eq34}
  \bar{\mathbf{G}}_e\left( \boldsymbol{r},\boldsymbol{r}^\prime \right) =-\mathrm{j}k\sum_{\alpha}{\mathbf{u}_{\alpha}^{\left( 4 \right)}\left( k\boldsymbol{r} \right) \mathbf{u}_{\alpha}^{\left( 1 \right)}\left( k\boldsymbol{r}^{\prime} \right)},r>r^\prime
\end{equation}
substituting this expression into \eqref{eq2} yields
\begin{equation}
  \boldsymbol{E}^s\left( \boldsymbol{r} \right) =-k^2\sum_{\alpha}{\left[ \eta _0\left< \mathbf{u}_{\alpha}^{\left( 1 \right)},\boldsymbol{J}^e \right> -\left< \mathbf{u}_{\bar{\alpha}}^{\left( 1 \right)},\mathrm{j}\boldsymbol{J}^m \right> \right] \mathbf{u}_{\alpha}^{\left( 4 \right)}\left( k\boldsymbol{r} \right)}.
\end{equation}
Comparing this result with \eqref{eq32}, we obtain
\begin{equation}
  \label{eq36}
  \mathbf{h}=-\mathbf{PI}
\end{equation}
where the projection operator
\begin{equation}
  \label{eq37}
  \mathbf{P}=\begin{bmatrix}
    \mathbf{P}^e&		-\mathbf{P}^m\\
  \end{bmatrix} 
\end{equation}
with components explicitly given by:
\begin{equation*}
  \left[ \mathbf{P}^e \right] _{\alpha i}=k\sqrt{\eta _0}\left< \mathbf{u}_{\alpha}^{\left( 1 \right)},\boldsymbol{\psi }_i \right> 
    \left[ \mathbf{P}^m \right] _{\alpha i}=k/\sqrt{\eta _0}\left< \mathbf{u}_{\bar{\alpha}}^{\left( 1 \right)},\boldsymbol{\psi }_i \right>.
\end{equation*}

\begin{figure}[!t]
  \centering
  \includegraphics[]{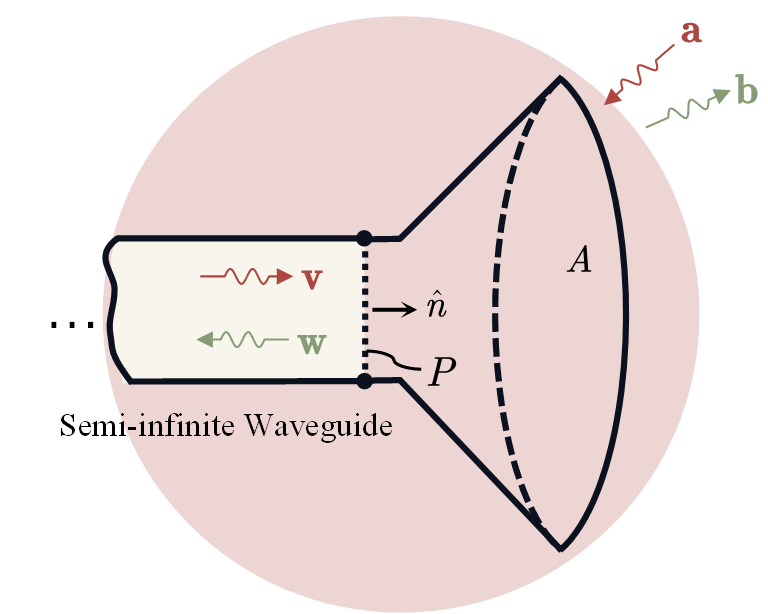}
  \caption{A system showcasing radiation, scattering, and guided waves. Here, $\mathbf{a}$ and $\mathbf{v}$ represent incoming waves, while $\mathbf{b}$ and $\mathbf{w}$ denote outgoing waves.}
  \label{fRadSca_Model}
\end{figure}

With the aforementioned notations, the GSM of an antenna can now be defined as
\begin{equation}
  \label{eq38}
  \begin{bmatrix}
    \mathbf{w}\\
    \mathbf{b}\\
  \end{bmatrix}=\tilde{\mathbf{S}}\begin{bmatrix}
    \mathbf{v}\\
    \mathbf{a}\\
  \end{bmatrix}
\end{equation}
where $\tilde{\mathbf{S}}$ explicitly consists of four distinct blocks:
\begin{equation}
  \tilde{\mathbf{S}}=\begin{bmatrix}
    \mathbf{\Gamma }&		\mathbf{R}\\
    \mathbf{T}&		\mathbf{S}\\
  \end{bmatrix}.
\end{equation}
Each block has a clear physical interpretation, linking the antenna's in-state vectors $\mathbf{v}, \mathbf{a}$ to out-state vectors $\mathbf{w}, \mathbf{b}$, as illustrated in Fig. \ref{fRadSca_Model}. Notably, our definition differs from that commonly used in \cite{ref_sph_near_measure}. Here, the submatrix $\mathbf{\Gamma}$ directly corresponds to the standard S-parameters of microwave circuit theory, whereas in \cite{ref_sph_near_measure}, it just represents reflection coefficient of the antenna, which is our special case for single-mode and single port feeding.

\subsection{Evaluating the GSM using MoM Data}

Here, we establish the relationship between the GSM and the MoM data. This relationship varies according to the integral equations formulation adopted. For the magnetic-type integral equations, substituting the expansions \eqref{eq32} of the incident fields $\boldsymbol{E}^{\mathrm{inc}}$ and $\boldsymbol{H}^{\mathrm{inc}}$ into the definition of $\mathbf{V}^{\mathrm{inc}}$ in \eqref{eq16}, and utilizing the projection operator $\mathbf{P}$ defined by \eqref{eq37}, we obtain
\begin{equation}
  \mathbf{V}^{\mathrm{inc}}=\mathbf{P}^t\mathbf{g}=2\mathbf{P}^t\mathbf{a}.
\end{equation}
Analogously, the waveguide excitation term $\mathbf{V}^{\mathrm{M}}$ becomes
\begin{equation}
  \mathbf{V}^{\mathrm{M}}=2\mathbf{Q}^{\mathrm{M}t}\mathbf{v}
\end{equation} 
therefore, 
\begin{equation}
  \mathbf{I}=\left( \mathbf{Z}^{\mathrm{M}} \right) ^{-1}\left( 2\mathbf{P}^t\mathbf{a}+2\mathbf{Q}^{\mathrm{M}t}\mathbf{v} \right).
\end{equation}
Then, \eqref{eq19}, \eqref{eq33} and \eqref{eq36} yield a compact relation between outgoing and incoming vectors
\begin{equation}
  \begin{split}
    &\mathbf{w}=-\left[ 2\mathbf{Q}^{\mathrm{M}}\left( \mathbf{Z}^{\mathrm{M}} \right) ^{-1}\mathbf{Q}^{\mathrm{M}t}+\mathbf{1} \right] \mathbf{v}-2\mathbf{Q}^{\mathrm{M}}\left( \mathbf{Z}^{\mathrm{M}} \right) ^{-1}\mathbf{P}^t\mathbf{a}\\
    &\mathbf{b}=-2\mathbf{P}\left( \mathbf{Z}^{\mathrm{M}} \right) ^{-1}\mathbf{Q}^{\mathrm{M}t}\mathbf{v}+\left[ \mathbf{1}-2\mathbf{P}\left( \mathbf{Z}^{\mathrm{M}} \right) ^{-1}\mathbf{P}^t \right] \mathbf{a}
  \end{split}
\end{equation}
with $\mathbf{1}$ being the identity matrix. Comparing this result with \eqref{eq38}, the GSM explicitly follows as:
\begin{equation*}
  \tilde{\mathbf{S}}=\begin{bmatrix}
    -2\mathbf{Q}^{\mathrm{M}}\left( \mathbf{Z}^{\mathrm{M}} \right) ^{-1}\mathbf{Q}^{\mathrm{M}t}-\mathbf{1}&		-2\mathbf{Q}^{\mathrm{M}}\left( \mathbf{Z}^{\mathrm{M}} \right) ^{-1}\mathbf{P}^t\\
    -2\mathbf{P}\left( \mathbf{Z}^{\mathrm{M}} \right) ^{-1}\mathbf{Q}^{\mathrm{M}t}&		-2\mathbf{P}\left( \mathbf{Z}^{\mathrm{M}} \right) ^{-1}\mathbf{P}^t+\mathbf{1}\\
  \end{bmatrix}.
\end{equation*}
By defining a combined projection operator $\tilde{\mathbf{P}}^{\mathrm{M}}$, this expression simplifies elegantly to
\begin{equation}
  \tilde{\mathbf{S}}=-2\tilde{\mathbf{P}}^{\mathrm{M}}\left( \mathbf{Z}^{\mathrm{M}} \right) ^{-1}\tilde{\mathbf{P}}^{\mathrm{M}t}+\tilde{\mathbf{1}}
\end{equation}
where
\begin{equation*}
  \tilde{\mathbf{P}}^{\mathrm{M}}=\begin{bmatrix}
    \mathbf{Q}^{\mathrm{M}}\\
    \mathbf{P}\\
  \end{bmatrix} ,\tilde{\mathbf{1}}=\begin{bmatrix}
    -\mathbf{1}_{M\times M}&		\\
    &		\mathbf{1}_{J\times J}\\
  \end{bmatrix}.
\end{equation*}

The evaluation of the GSM via electric-type integral equations proceeds similarly. Here we directly present the corresponding result as:
\begin{equation}
  \label{eq45}
  \tilde{\mathbf{S}}=-2\tilde{\mathbf{P}}^{\mathrm{E}}\left( \mathbf{Z}^{\mathrm{E}} \right) ^{-1}\tilde{\mathbf{P}}^{\mathrm{E}t}+\mathbf{1}.
\end{equation} 
Note here $\mathbf{1}$ is the standard identity matrix and the projection matrix $\tilde{\mathbf{P}}^{\mathrm{E}}$ is defined by
\begin{equation}
  \tilde{\mathbf{P}}^{\mathrm{E}}=\begin{bmatrix}
    \mathbf{Q}^{\mathrm{E}}\\
    \mathbf{P}\\
  \end{bmatrix}.
\end{equation}

In practical computations, scattering coefficients associated with waveguide evanescent modes are typically irrelevant. Hence, the corresponding rows in $\tilde{\mathbf{P}}^{\mathrm{M}}$ and $\tilde{\mathbf{P}}^{\mathrm{E}}$ may be omitted. Under these conditions, the GSM satisfies unitarity for lossless systems, \textit{i.e.}, $\tilde{\mathbf{S}}^{\mathrm{H}}\tilde{\mathbf{S}}=\mathbf{1}$, providing a robust verification criterion for the correctness of the numerical GSM evaluation. Here, $(\cdot)^\mathrm{H}$ denotes the conjugate transpose.

\section{Compression Storage of the GSM}
\label{Sec_IV}

In typical applications, storing GSM data for all frequencies and antennas is usually necessary. Although this is much easier than directly storing the MoM impedance matrix, it still presents a significant storage burden. However, radiation and scattering problems are primarily influenced by the leading components of the GSM, allowing for effective compression and storage.

In lossless cases, the unitarity of $\tilde{\mathbf{S}}$ enables unitary diagonalization based on eigenvalue decomposition:
\begin{equation}
  \label{eq47}
  \tilde{\mathbf{S}}\begin{bmatrix}
    \mathbf{w}_n\\
    \mathbf{b}_n\\
  \end{bmatrix} =s_n\begin{bmatrix}
    \mathbf{w}_n\\
    \mathbf{b}_n\\
  \end{bmatrix}
\end{equation}
For simplicity, we define $\tilde{\mathbf{f}}_n = \left[ \mathbf{w}_n^t, \mathbf{b}_n^t \right]^t$. Physically, equation \eqref{eq47} indicates that when the system is excited by an in-state eigenvector $\tilde{\mathbf{a}}_n = s_{n}^{-1}\tilde{\mathbf{f}}_n$, the resulting outgoing state is simply a scaled version of itself, multiplied by the eigenvalue $s_n$ ($\left| s_n \right|= 1$). The most significant scattering contributions correspond to eigenvectors $\tilde{\mathbf{f}}_n$ whose eigenvalues $s_n$ approach -1. Therefore, by storing only the first $N$ dominant eigenmodes, we achieve effective compression and storage of the GSM.

However, the unitary nature of the GSM leads to numerical convergence difficulties in direct eigenvalue decomposition. To overcome this, it is advantageous to instead decompose the shifted matrix $\tilde{\mathbf{T}} = (\tilde{\mathbf{S}} - \mathbf{1}) / 2$. It can readily be shown that $\tilde{\mathbf{T}}$ shares identical eigenvectors with $\tilde{\mathbf{S}}$, while their eigenvalues are related through $t_n = (s_n - 1) / 2$. Thus, the dominant scattering modes correspond to eigenvectors with eigenvalues $t_n$ close to -1, equivalently those with largest modulus. The GSM can then be efficiently reconstructed using these dominant eigenvectors:
\begin{equation}
  \label{eq48}
  \tilde{\mathbf{S}}\simeq \mathbf{1}+2\tilde{\mathbf{F}} \begin{bmatrix}
    t_1&		&		\\
    &		\ddots&		\\
    &		&		t_N\\
  \end{bmatrix} \tilde{\mathbf{F}}^{\mathrm{H}}
\end{equation}
where $\tilde{\mathbf{F}} = \left[ \begin{matrix} \tilde{\mathbf{f}}_1 & \cdots & \tilde{\mathbf{f}}_N \end{matrix} \right]$, and $N$ is chosen as the largest $n$ that $\left| t_n \right| > \iota \left| t_1 \right|$. The threshold parameter $\iota$, empirically set to $\iota = 1.53\times 10^{-5}$, is further discussed in Sec. \ref{Sec_V_C}.

The eigenvalue decomposition of $\tilde{\mathbf{T}}$ can be connected to the characteristic modes defined by the impedance matrix $\mathbf{Z}^{\mathrm{E}}$, obtained from electric-type integral equations:
\begin{equation}
  \label{eq49}
  \mathbf{Z}^{\mathrm{E}}\mathbf{I}_n=\left( 1+\mathrm{j}\lambda _n \right) \mathbf{R}^{\mathrm{E}}\mathbf{I}_n \text{ or } \mathbf{X}^{\mathrm{E}}\mathbf{I}_n=\lambda _n\mathbf{R}^{\mathrm{E}}\mathbf{I}_n
\end{equation}
where $\mathbf{R}^{\mathrm{E}}$ and $\mathbf{X}^{\mathrm{E}}$ denote the real and imaginary parts of $\mathbf{Z}^{\mathrm{E}}$.

Because the difference between the outgoing vector and incoming vector is [from \eqref{eq28}, \eqref{eq33} and \eqref{eq36}]:
\begin{equation}
  \begin{bmatrix}
    \mathbf{w}\\
    \mathbf{b}\\
  \end{bmatrix} -\begin{bmatrix}
    \mathbf{v}\\
    \mathbf{a}\\
  \end{bmatrix} =-\tilde{\mathbf{P}}^{\mathrm{E}}\mathbf{I}
\end{equation}
thus, for eigenvectors associated with the GSM, we have: 
\begin{equation}
  \label{eq51}
  -\tilde{\mathbf{P}}^{\mathrm{E}}\tilde{\mathbf{I}}_n=\tilde{\mathbf{f}}_n-\tilde{\mathbf{a}}_n=\left( 1-s_{n}^{-1} \right) \tilde{\mathbf{f}}_n.
\end{equation}

Noting the identity $\mathbf{R}^{\mathrm{E}}=\tilde{\mathbf{P}}^{\mathrm{E}t}\tilde{\mathbf{P}}^{\mathrm{E}}$, we rearrange the left-hand side equation in \eqref{eq49} as
\begin{equation}
  -\frac{1}{1+\mathrm{j}\lambda _n}\mathbf{I}_n=\left( \mathbf{Z}^{\mathrm{E}} \right) ^{-1}\tilde{\mathbf{P}}^{\mathrm{E}t}\left( 1-s_{n}^{-1} \right) \tilde{\mathbf{f}}_n.
\end{equation}
Multiplying this equation by $-\tilde{\mathbf{P}}^{\mathrm{E}}$ from the left, substituting the result from \eqref{eq51}, and cancelling the common scalar $\left( 1-s_{n}^{-1} \right) $, yields:
\begin{equation}
  -\frac{1}{1+\mathrm{j}\lambda _n}\tilde{\mathbf{f}}_n=-\tilde{\mathbf{P}}^{\mathrm{E}}\left( \mathbf{Z}^{\mathrm{E}} \right) ^{-1}\tilde{\mathbf{P}}^{\mathrm{E}t}\tilde{\mathbf{f}}_n
\end{equation}

Recalling from \eqref{eq45} that $\tilde{\mathbf{T}}=-\tilde{\mathbf{P}}^{\mathrm{E}}\left( \mathbf{Z}^{\mathrm{E}} \right) ^{-1}\tilde{\mathbf{P}}^{\mathrm{E}t}$, we conclude the relationship $t_n=-\frac{1}{1+\mathrm{j}\lambda _n}$. The special condition $t_n=-1$ corresponds exactly to
\begin{equation}
  \lambda _n=\frac{\mathbf{I}_{n}^{\mathrm{H}}\mathbf{X}^{\mathrm{E}}\mathbf{I}_n}{\mathbf{I}_{n}^{\mathrm{H}}\mathbf{R}^{\mathrm{E}}\mathbf{I}_n}=0
\end{equation}
indicating zero net reactive power (numerator) and thus representing a scattering resonance (external resonance) \cite{ref_CMA_Harrington}. This clarifies why the dominant GSM components typically possess eigenvalues $t_n$ near -1. Note here we have used the right definition in \eqref{eq49} for expressing $\lambda_n$. 

In lossy systems, eigenvectors lose orthogonality, diminishing the effectiveness of eigenvalue-based compression procedure like \eqref{eq48}. In these situations, singular value decomposition (SVD) of $\tilde{\mathbf{T}}$ provides a robust alternative. Notably, the singular value $\sigma_n=\left| t_n \right|^2\le 1$, and $\sigma_n=1$ corresponding to $t_n = -1$, thus effectively capturing dominant scattering modes.

\section{Numerical Results}
\label{Sec_V}

\subsection{Validation example}
\label{Sec_V_A}

This section validates the GSM formulation derived in previous sections. Since GSM computation is not yet incorporated into mainstream electromagnetic simulation software, direct verification is limited mainly to the $\mathbf{\Gamma}$ sub-block of $\tilde{\mathbf{S}}$. Other sub-blocks require indirect verification. For instance, when exciting an antenna solely through waveguide ports, the $\mathbf{T}$ sub-block relates waveguide mode coefficients to the spherical wave expansion of radiated fields, allowing verification through radiation pattern and gain comparisons. Similarly, when no waveguide mode excitation is present, the $\mathbf{S}$ sub-block connects incident plane waves to scattered fields, enabling verification through radar cross-section (RCS) calculations. Owing to transmit/receive reciprocity, $\mathbf{R}=\mathbf{T}^t$ requires no additional verification.

\begin{figure}[!t]
  \centering
  \includegraphics[]{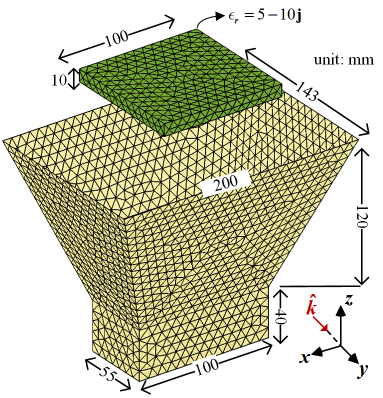}
  \caption{A multimode horn antenna with a lossy cuboid positioned above it.}
  \label{fHorn_illuDieModel}
\end{figure}

As an illustrative validation case, we analyze a multimode horn antenna with an adjacent lossy cuboid, as shown in Fig. \ref{fHorn_illuDieModel}. The electromagnetic currents on the geometry are represented by 7783 RWG basis functions. The external fields are expanded into $J=1056$ spherical wavefunctions. Given the rectangular waveguide feed dimensions, the frequency range from 3.2 to 3.8 GHz supports five propagating modes, resulting in a GSM of dimensions $1061\times1061$.

\begin{figure}[!t]
  \centering
  \includegraphics[]{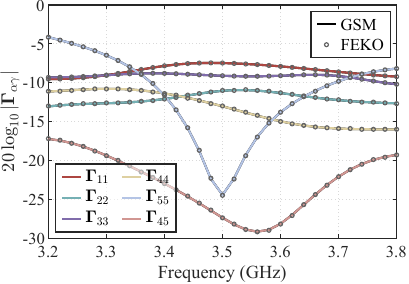}
  \caption{S-parameters among different feeding modes. Modes 1 to 5 represent TE$_{10}$, TE$_{20}$, TE$_{01}$, TE$_{11}$ and TM$_{11}$ modes, respectively.}
  \label{f_SparaHorn}
\end{figure}

The top-left $5\times 5$ submatrix of $\tilde{\mathbf{S}}$, \textit{i.e.}, $\mathbf{\Gamma}$, corresponds directly to the S-parameters between the five waveguide modes. Fig. \ref{f_SparaHorn} illustrates these S-parameters over frequency, exhibiting excellent agreement with results obtained from FEKO simulations.

\begin{figure}[!t]
  \centering
  \subfloat[]{\includegraphics[]{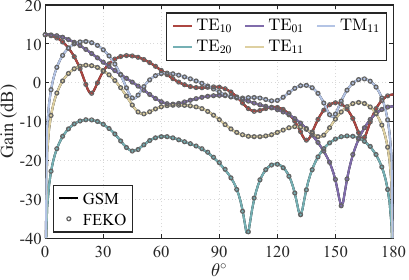}%
  \label{f_GainHorn_E}}
  \vfil
  \subfloat[]{\includegraphics[]{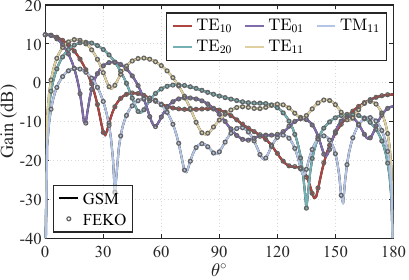}%
  \label{f_GainHorn_H}}
  \caption{Gain pattern at 3.5 GHz. (a) xoz-plane, (b) yoz-plane.}
\label{f_GainHorn}
\end{figure}

To indirectly verify the accuracy of the $\mathbf{T}$ matrix, we excite only the waveguide modes (thus, $\mathbf{a}=\mathbf{0}$). Under this condition, the outgoing spherical-wave vector is simply $\mathbf{b}=\mathbf{T}\mathbf{v}$. The corresponding antenna gain patterns computed from this relationship are presented in Fig. \ref{f_GainHorn}. Each curve corresponds to independently exciting one of the five waveguide eigenmodes. Again, our GSM-based results closely match FEKO's solutions (indicated by circles), confirming the correctness of the $\mathbf{T}$ sub-block and thus also the 
$\mathbf{R}$ sub-block.

\begin{figure}[!t]
  \centering
  \includegraphics[]{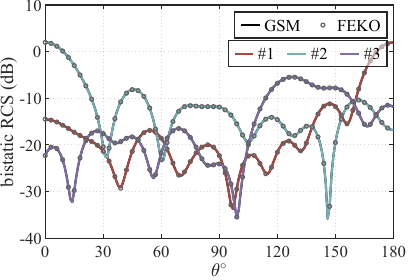}
  \caption{RCS at $\phi=0$ plane. This figure shows RCS results in three scenarios where the plane wave directions are $\hat{k}_i=-\hat z,-\hat x,\hat z$, all with polarization angle 0$^\circ$.}
  \label{fRCS_Horn}
\end{figure}

When illuminating the antenna structure with a plane wave ($\mathbf{v}=\mathbf{0}$), the scattered spherical waves are described by the expansion vector $\mathbf{h}=\mathbf{b}-\mathbf{a}$, with $\mathbf{b}=\mathbf{S}\mathbf{a}$. Appendix~\ref{APP_A} provides the spherical-wave coefficients corresponding to the plane-wave incidence. Fig. \ref{fRCS_Horn} shows the bistatic RCS results for three representative incidence directions. These GSM-derived results again align well with FEKO simulations. Although exhaustive validation for every incident direction and polarization is impractical, the presented cases sufficiently verify the accuracy of the $\mathbf{S}$ sub-block.

\subsection{Application example}
\label{Sec_V_B}

The previous subsection primarily validated the GSM formulation, without emphasizing practical applications. Here, we illustrate GSM's significant potential to accelerate antenna array simulations. To enhance clarity and facilitate the subsequent discussion, we briefly repeat essential mathematical relations, noting that some details can also be found in \cite{ref_Smat_FEM}.

Consider an antenna array where the GSM of the $p$-th array element is expressed as
\begin{equation}
  \label{eq55}
  \begin{bmatrix}
    \mathbf{\Gamma }^p&		\mathbf{R}^p\\
    \mathbf{T}^p&		\mathbf{S}^p-\mathbf{1}\\
  \end{bmatrix}  \begin{bmatrix}
    \mathbf{v}^p\\
    \mathbf{a}^p\\
  \end{bmatrix} =\begin{bmatrix}
    \mathbf{w}^p\\
    \mathbf{h}^p\\
  \end{bmatrix}.
\end{equation}
We assume that the array excitation is provided exclusively through the antenna ports. Thus, for the $p$-th antenna, the incoming spherical wave vector $\mathbf{a}^p$ comprises scattered fields ($\mathbf{h}^q$) from all other antennas ($q\ne p$). Employing the translation property of spherical waves, we write:
\begin{equation}
  \label{eq56}
  \mathbf{a}^p=\sum_{q\ne p} \bm{\mathcal{G}}_{pq}\mathbf{h}^q
\end{equation}
where $\bm{\mathcal{G}}_{pq}=\bm{\mathcal{Y}}\left(k\boldsymbol{d}_{pq}\right)/2$ is the spherical-wave translation matrix, and $\boldsymbol{d}_{pq}$ denotes the vector distance from the center of antenna $q$ to antenna $p$. For detailed analytical formulations of matrix $\bm{\mathcal{Y}}$, cf. \cite[Appendix C]{ref_my}. In scenarios where the distance $d_{pq}$ is less than the minimal enclosing radius $r_{\min}$, the analytical formulation may not strictly apply, and a numerical calculation of $\bm{\mathcal{G}}_{pq}$ based on \cite{ref_translation_PRA} becomes necessary.

Substituting \eqref{eq56} into \eqref{eq55}, we obtain the following system of equations
\begin{equation}
\begin{cases}
	\displaystyle\mathbf{\Gamma }^p\mathbf{v}^p+\mathbf{R}^p\sum_{q\ne p}{\bm{\mathcal{G}} _{pq}\mathbf{h}^q}=\mathbf{w}^p\\
	\displaystyle\mathbf{T}^p\mathbf{v}^p+\left( \mathbf{S}^p-\mathbf{1} \right) \sum_{q\ne p}{\bm{\mathcal{G}} _{pq}\mathbf{h}^q}=\mathbf{h}^p.\\
\end{cases}
\end{equation}
Collecting these equations across all antennas yields a matrix form:
\begin{equation}
\label{eq58}
\begin{cases}
    \hat{\mathbf{\Gamma}}\mathbf{v}+\hat{\mathbf{R}}\hat{\bm{\mathcal{G}}}\hat{\mathbf{h}}=\mathbf{w}\\
    \hat{\mathbf{T}}\mathbf{v}+\left( \hat{\mathbf{S}}-\mathbf{1} \right) \hat{\bm{\mathcal{G}}}\hat{\mathbf{h}}=\hat{\mathbf{h}}\\
\end{cases}
\end{equation}
with definitions
\begin{equation*}
    \hat{\bm{\mathcal{G}}}=\begin{bmatrix}
	\mathbf{0}&		\bm{\mathcal{G}}_{12}&		\cdots\\
	\bm{\mathcal{G}} _{21}&		\mathbf{0}&		\vdots\\
	\vdots&		\cdots&		\ddots\\
\end{bmatrix}
\end{equation*}
$\hat{\mathbf{\Gamma}}$, $\hat{\mathbf{R}}$, $\hat{\mathbf{T}}$, $\hat{\mathbf{S}}$ are block diagonal matrices, where $\hat{\mathbf{\Gamma}}=\mathrm{diag}(\mathbf{\Gamma}^1,\mathbf{\Gamma}^2,\cdots)$, and so forth.
\begin{equation*}
  \mathbf{w}=\begin{bmatrix}
    \mathbf{w}^1\\
    \mathbf{w}^2\\
    \vdots\\
  \end{bmatrix} ,\mathbf{v}=\begin{bmatrix}
    \mathbf{v}^1\\
    \mathbf{v}^2\\
    \vdots\\
  \end{bmatrix} ,\hat{\mathbf{h}}=\begin{bmatrix}
    \mathbf{h}^1\\
    \mathbf{h}^2\\
    \vdots\\
  \end{bmatrix}
\end{equation*}

Solving for $\hat{\mathbf{h}}$ from the second equation in \eqref{eq58} and inserting into the first yields
\begin{equation}
    \label{eq59}
    \left\{ \hat{\mathbf{\Gamma}}+\hat{\mathbf{R}}\hat{\bm{\mathcal{G}}}\left[ \mathbf{1}-\left( \hat{\mathbf{S}}-\mathbf{1} \right) \hat{\bm{\mathcal{G}}} \right] ^{-1}\hat{\mathbf{T}} \right\} \mathbf{v}=\mathbf{w}.
\end{equation}
Thus, the synthesized S-parameters among antenna array elements are expressed as:
\begin{equation}
    \label{eq60}
    \mathbf{\Gamma}=\hat{\mathbf{\Gamma}}+\hat{\mathbf{R}}\hat{\bm{\mathcal{G}}}\left[ \mathbf{1}-\left( \hat{\mathbf{S}}-\mathbf{1} \right) \hat{\bm{\mathcal{G}}} \right] ^{-1}\hat{\mathbf{T}}.
\end{equation}

\begin{figure}[!t]
  \centering
  \includegraphics[]{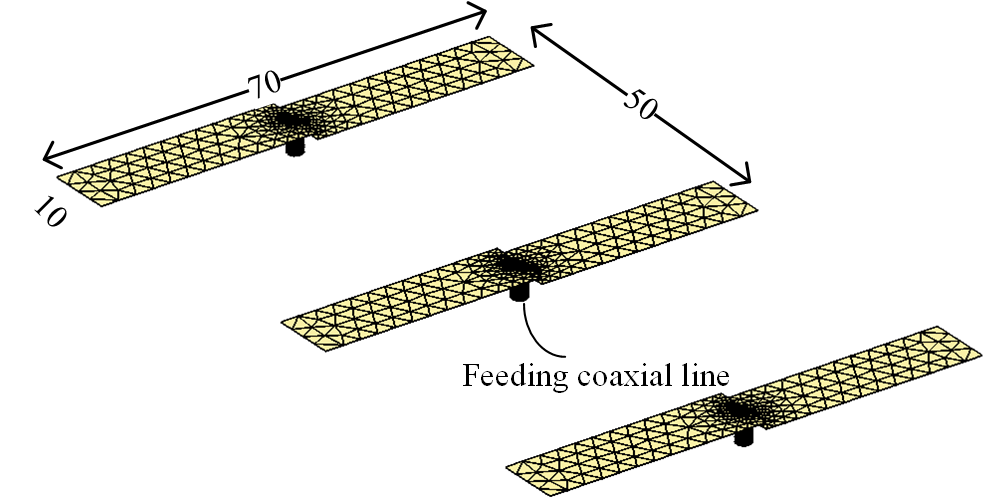}
  \caption{Geometry of the three-element dipole array. Each dipole is fed by a 50$\Omega$ coaxial waveguide (inner diameter: 0.5 mm, outer diameter: 1.15 mm). All dimensions are given in millimeters.}
  \label{fDipoleArray_Model}
\end{figure}

\begin{figure}[!t]
  \centering
  \includegraphics[]{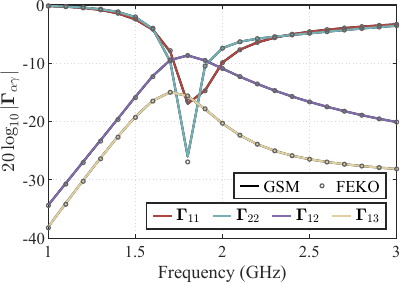}
  \caption{S-parameters of the 3-element dipole array: GSM-based results (solid lines) compared to full-wave FEKO simulations (circles).}
  \label{fDipoleArray3_Spara}
\end{figure}

To demonstrate the developed formulation, we consider a 3-element dipole array fed via coaxial waveguides, as depicted in Fig.~\ref{fDipoleArray_Model}. Initially, we calculate the GSM for a single dipole element, and then calculate the complete array S-parameters via \eqref{eq60}. Fig. \ref{fDipoleArray3_Spara} shows these synthesized S-parameters across the frequency range of 1-3 GHz, exhibiting excellent agreement with full-wave simulations of the entire array performed in FEKO.

The example is analyzed across 21 frequency points from 1 GHz to 3 GHz at intervals of 100 MHz. Each antenna element utilizes 2526 RWG basis functions and 646 spherical vector wavefunctions. The GSM-based approach outlined in \eqref{eq60} required a total computation time of approximately 116 seconds (with 103 seconds dedicated to computing the antenna element's GSM), less than the 294 seconds taken by full-wave FEKO simulations (employing four parallel threads). While the computational benefit of GSM may seem modest for this single calculation, its advantages become evident when multiple array configurations with identical antenna elements are analyzed. Specifically, changes in the array layout only require recalculating the translation matrix $\hat{\bm{\mathcal{G}}}$, which takes around 13 seconds in total. 

The main computational cost in \eqref{eq60} arises from inverting the matrix $\left[ \mathbf{1}-\left( \hat{\mathbf{S}}-\mathbf{1} \right) \hat{\bm{\mathcal{G}}} \right]^{-1}$, whose complexity scales quadratically with the number of array elements. This behavior mirrors direct MoM solutions of entire arrays; though, GSM is often more computationally efficient due to the smaller number of spherical wavefunctions relative to RWG basis functions per antenna element.

To further alleviate computational burdens, we present an iterative solution, based on \eqref{eq58}, as follows:
\begin{equation}
  \hat{\mathbf{h}}\left( l+1 \right) =\left( \hat{\mathbf{S}}-\mathbf{1} \right) \tilde{\bm{\mathcal{G}}}\hat{\mathbf{h}}\left( l \right) 
\end{equation}
initialized by $\hat{\mathbf{h}}\left( 0 \right) =\hat{\mathbf{T}}\mathbf{v}$. Then $\hat{\mathbf{h}}$ is solved by the summation of $\hat{\mathbf{h}}\left( l \right)$ from $l=0$. Typically, convergence is reached within a dozen iterations, significantly accelerating computations since each iteration involves only matrix-vector multiplications.

\begin{figure}[!t]
  \centering
  \includegraphics[]{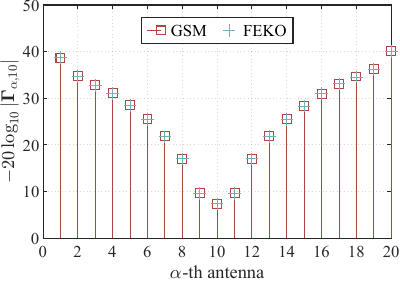}
  \caption{S-parameters among the tenth dipole and other dipoles in a 20-element dipole array. The FEKO results is obtained using MLFMM.}
  \label{fSpara_DipoleArray20}
\end{figure}

To illustrate this benefit concretely, we consider a 20-element dipole array (geometry analogous to Fig. \ref{fDipoleArray_Model}). The direct GSM calculation, based on \eqref{eq60}, required 205 seconds, whereas the iterative procedure reduced the total computation time significantly to just 52 seconds. Of these times, 35 seconds were dedicated to computing the translation matrix $\tilde{\bm{\mathcal{G}}}$. For further illustration, we compared the iterative results against FEKO's multilevel fast multipole method (MLFMM) simulations. Fig.~\ref{fSpara_DipoleArray20} displays the S-parameters between the tenth dipole and the other elements at 1.9 GHz. The iterative results exhibit excellent agreement with those obtained via MLFMM. However, it is noteworthy that the MLFMM computations required approximately 451 seconds per frequency point, with four parallel computational threads. Although a direct comparison to FEKO's conventional MoM implementation was not conducted, it is expected that the computational cost for MoM would be significantly higher than MLFMM, further highlighting the efficiency and practical value of the GSM-based iterative procedure presented herein.

\subsection{Threshold $\iota$ for the Compression Storage Method}
\label{Sec_V_C}

\begin{figure}[!t]
  \centering
  \subfloat[]{\includegraphics[]{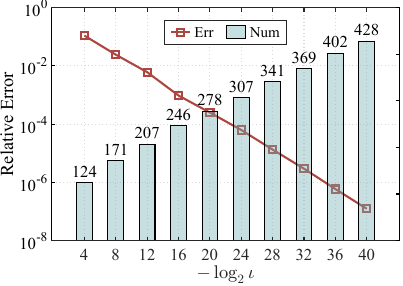}%
  \label{f_err_Horn}}
  \vfil
  \subfloat[]{\includegraphics[]{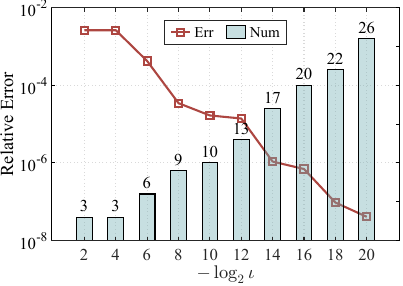}%
  \label{f_err_dipole}}
  \caption{Threshold, relative Error, and number of leading components. The line represents the relative error, and the bars represent the number of leading components. (a) for the structure in Fig. \ref{fHorn_illuDieModel}, (b) for the dipole element in Fig. \ref{fDipoleArray_Model}.}
\label{f_err}
\end{figure}

In the final example, we investigates the accuracy of the compressed GSM ($\tilde{\mathbf{S}}^{\prime}$) under varying threshold values of $\iota$. We define the average relative error between the compressed and original GSM as follows:
\begin{equation}
  \mathrm{Err}=\frac{1}{N}\sum_n^N{\frac{\left\| \tilde{\mathbf{S}}\tilde{\mathbf{f}}_{n}^{\mathrm{in}}-\tilde{\mathbf{S}}^{\prime}\tilde{\mathbf{f}}_{n}^{\mathrm{in}} \right\| _2}{\left\| \tilde{\mathbf{S}}\tilde{\mathbf{f}}_{n}^{\mathrm{in}} \right\| _2}}
\end{equation}
where $\tilde{\mathbf{f}}_{n}^{\mathrm{in}}$ are randomly generated in-state vectors, and the total number of test vectors is set to $N=100$. Fig. \ref{f_err}a presents the relative reconstruction error versus the threshold parameter $-\log _2\iota $ for the GSM associated with the multimode horn antenna structure in Fig.\ref{fHorn_illuDieModel}. The error decreases monotonically as the threshold becomes stricter ($-\log _2\iota $ increases). At $-\log _2\iota = 16$, the error falls below $10^{-3}$, requiring retention of only the 246 leading eigenmodes. Thus, compressed GSM storage necessitates approximately $1061\times246\times2$ complex numbers, achieving a memory saving of about 53.7\%. Furthermore, for lossless cases, memory savings could potentially be doubled.

For comparison, Fig. \ref{f_err}b depicts the compression performance for the GSM of the narrowband dipole antenna element shown in Fig. \ref{fDipoleArray_Model}. Here, the error decreases rapidly with increasing $-\log _2\iota $. At just $-\log _2\iota = 6$, the relative reconstruction error already falls below $10^{-3}$, requiring retention of merely 6 leading modes. Consequently, compressed storage demands just $646\times6$ complex numbers, representing a substantial memory saving of approximately 99.1\%.

Based on these numerical experiments, we recommend using a threshold parameter $-\log _2\iota = 16$ (\textit{i.e.}, $\iota \approx 1.53\times 10^{-5}$) as an optimal compromise, offering robust GSM accuracy combined with significant memory savings in practical antenna applications.

\section{Conclusion}
This paper presented a concise and robust method for computing the GSM of antennas via MoM. Two integral equation formulations (magnetic-type and electric-type) were developed, leading to explicit algebraic expressions linking GSM directly with MoM impedance matrices. Numerical validations, exemplified by a multimode horn antenna, confirmed the accuracy and effectiveness of the proposed formulations.

We also demonstrated that GSM data can be efficiently compressed, significantly alleviating memory requirements. Specifically, eigenvalue decomposition was utilized in lossless scenarios, whereas SVD was effectively employed for lossy conditions. A clear theoretical connection with characteristic mode theory provided valuable insights into the mechanisms underlying GSM compression.

Moreover, an efficient iterative algorithm was proposed for antenna array analysis using GSM, substantially reducing computational times compared with conventional methods such as the MLFMM. The practicality and efficiency highlighted in the simulation of a 20-element dipole array demonstrate GSM's great potential for accelerating antenna array designs.

These contributions collectively encourage further exploration and broader application of GSM-based methods in computational electromagnetics, particularly for rapid analysis and design optimization of complex antenna arrays.

\begin{appendices}

  \section{Expanding Uniform Plane Waves via Vector Spherical Wavefunctions}
  \label{APP_A}
  
  Consider a uniform plane wave represented by $\boldsymbol{E}^\mathrm{inc}=\boldsymbol{E}_0e^{-\mathrm{j}k\hat{k}_i\cdot \boldsymbol{r}}$, where $\boldsymbol{E}_0$ is the amplitude vector, and $\hat{k}_i$ is the unit wave vector in the direction of propagation. The expansion coefficients under regular vector spherical waves are:
  \begin{equation}
    g_{\alpha}=\frac{4\pi}{k\sqrt{\eta _0}}\left[ \mathrm{j}^{-l}\delta _{\tau 1}-\mathrm{j}^{-\left( l+1 \right)}\delta _{\tau 2} \right] \boldsymbol{A}_{\tau \alpha}\left( \hat{k}_i \right) \cdot \boldsymbol{E}_0
  \end{equation}
where the vector spherical harmonics $\boldsymbol{A}_{\tau\alpha}$ are given in \cite{ref_scattering_theory,ref_hybrid}. Note that the incoming components $\mathbf{a}$ is half of $\mathbf{g}$.

\end{appendices}

  \begin{IEEEbiography}[{\includegraphics[width=1in,height=1.25in,clip,keepaspectratio]{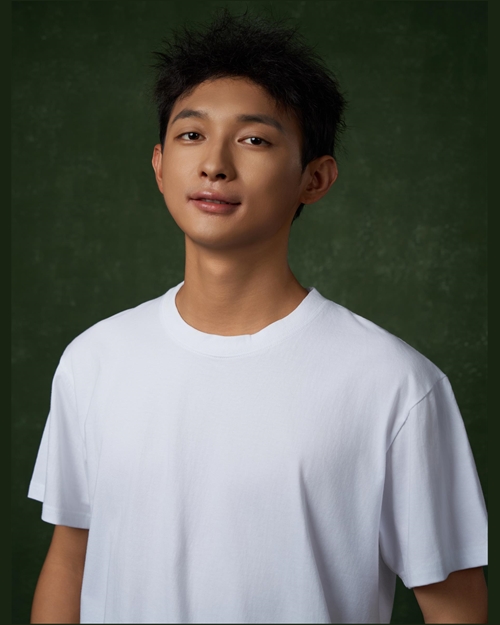}}]{Chenbo Shi}
  Chenbo Shi was born in 2000 in China. He received his Bachelor's degree from the University of Electronic Science and Technology of China (UESTC) in 2022. He is currently pursuing his Ph.D. at the same institution. His research interests include electromagnetic theory, characteristic mode theory, and computational electromagnetics. 
  
  Chenbo has been actively involved in several research projects and has contributed to publications in these areas. His work aims to advance the understanding and application of electromagnetic phenomena in various technological fields.
  \end{IEEEbiography}
    
  \begin{IEEEbiography}[{\includegraphics[width=1in,height=1.25in,clip,keepaspectratio]{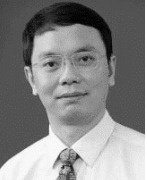}}]{Jin Pan}
    received the B.S. degree in electronics and communication engineering from the Radio Engineering Department, Sichuan University, Chengdu, China, in 1983, and the M.S. and Ph.D. degrees in electromagnetic field and microwave technique from the University of Electronic Science and Technology of China (UESTC), Chengdu, in 1983 and 1986, respectively. 
    
    From 2000 to 2001, he was a Visiting Scholar in electronics and communication engineering with the Radio Engineering Department, City University of Hong Kong. He is currently a Full Professor with the School of Electronic Engineering, UESTC. 
    
    His current research interests include electromagnetic theories and computations, antenna theories, and techniques, field and wave in inhomogeneous media, and microwave remote sensing theories and its applications. 
  \end{IEEEbiography}

  \begin{IEEEbiography}[{\includegraphics[width=1in,height=1.25in,clip,keepaspectratio]{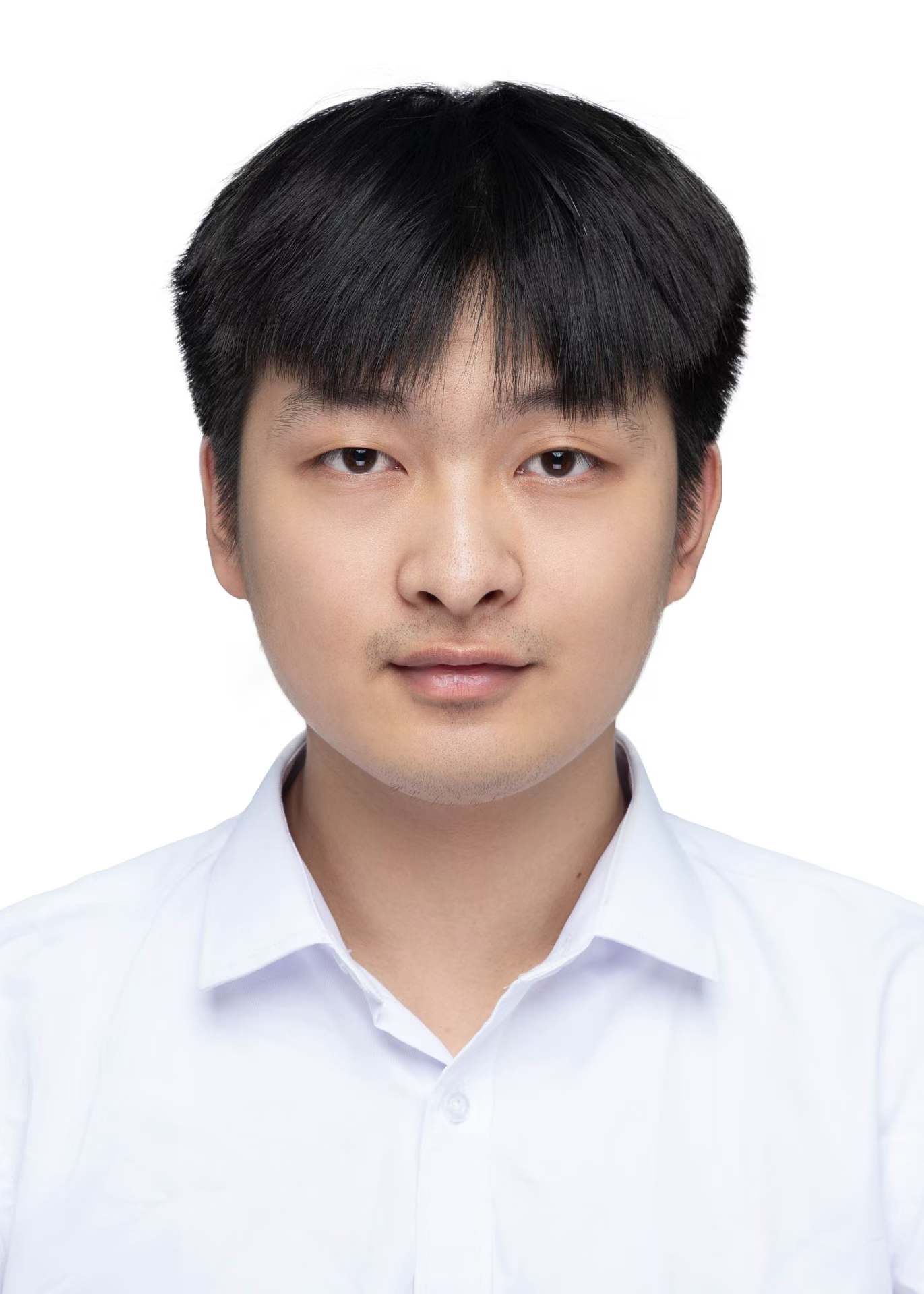}}]{Xin Gu}
  received the B.E. degree from Chongqing University of Posts and Telecommunications (CQUPT), ChongQing, China, in 2022. He is currently pursuing the M.S. degree with the School of Electronic Science and Engineering, University of Electronic Science and Technology of China (UESTC), Chengdu, China.
    
  His research interests include electromagnetic theory and electromagnetic measurement techniques
  \end{IEEEbiography}
  
  \begin{IEEEbiography}[{\includegraphics[width=1in,height=1.25in,clip,keepaspectratio]{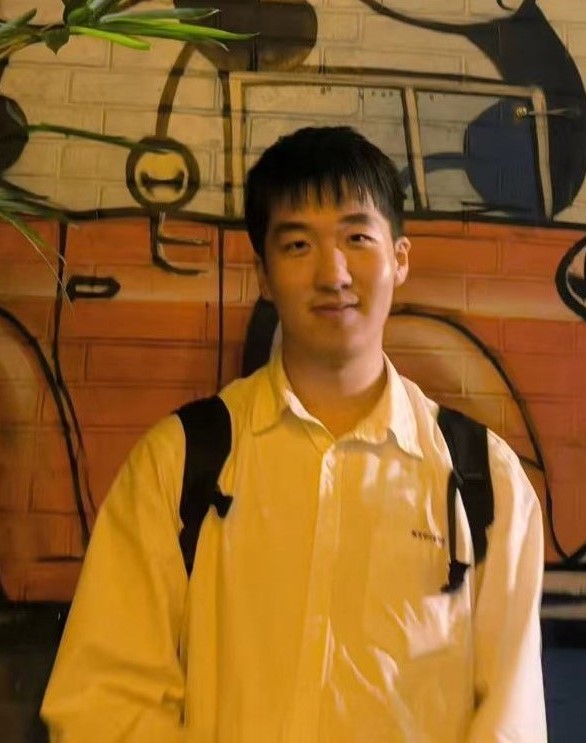}}]{Shichen Liang}
  received the B.E. degree from Beijing University of Chemical Technology (BUCT), Beijing, China, in 2022. He is currently pursuing the M.S. degree with the School of Electronic Science and Engineering, University of Electronic Science and Technology of China (UESTC), Chengdu, China. 
  
  His research interests include electromagnetic theory and electromagnetic measurement techniques.
  \end{IEEEbiography}

  \begin{IEEEbiography}[{\includegraphics[width=1in,height=1.25in,clip,keepaspectratio]{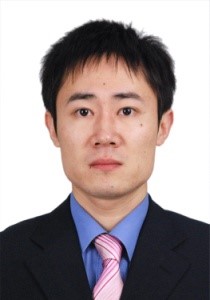}}]{Le Zuo}
  received the B.Eng., M.Eng. and Ph.D. degrees in electromagnetic field and microwave techniques from the University of Electronic Science and Technology of China (UESTC), in 2004, 2007 and 2018, respectively. 
  
  From 2017 to 2018, he was a Research Associate with the School of Electrical and Electronic Engineering, Nanyang Technological University, Singapore. He is currently a Research Fellow with the 29th Institute of the China Electronics Technology Group, Chengdu, China. His research interests include antenna theory and applications.
  \end{IEEEbiography}

\end{document}